\documentclass[aps,physrev,preprint,groupedaddress]{revtex4-2}

\usepackage[english]{babel}
\usepackage{graphicx} % inclusion des figures
\usepackage{caption}
\usepackage{subcaption}
\usepackage[separate-uncertainty=true]{siunitx}
\usepackage{physics}
\usepackage[dvipsnames]{xcolor}
\newcommand\Rey{\mbox{\textit{Re}}}

\begin{document}

%\preprint{}

\title{Structural and mechanical characterisation of nylon fiber aggregates.}


\author{L. Gey}
\author{R. Guenin}
\author{P. Le Gal}
\author{G. Verhille}
\email[L.Gey mail :]{lucas.gey@univ-amu.fr}
%\thanks{}
\affiliation{Aix Marseille Univ., CNRS, Centrale Méditerranée, IRPHE, F-13013 Marseille, France}


\date{\today}

\begin{abstract}
The study of fiber networks is of great importance due to their presence in a wide range of natural and industrial processes. 
These networks are particularly complex because their properties depend not only on the characteristics of individual fibers but also on their collective arrangements within the network.
In this study, we investigate the case of nylon fiber aggregates generated in a turbulent von Kármán  flow.
The key properties of these aggregates are characterized, including their three-dimensional structures, analyzed through X-ray tomography, and their mechanical responses. 
The results are compared with those observed for aegagropilae, a natural fiber aggregate commonly found on Mediterranean seashores
\end{abstract}

%\keywords{}

\maketitle

\section{\label{SecIntro} Introduction}

A wide range of materials consists of ensembles of elongated objects. 
The biological world offers numerous examples, with filaments appearing at various scales and in diverse organizations, serving a wide range of functions, from macromolecular networks of cytoskeleton at the scale of a cell to centimeter long collagen aggregates in tendons~\cite{picu_mechanics_2011}.
Structures formed from elongated objects can also be man-made. For instance, nonwovens, composed of long entangled fibers, can be used for thermal or sound insulation~\cite{karimi_acoustic_2022} and filtration~\cite{hutten_handbook_2007}. Similarly, the paper industry relies on cellulose fibers to create strong yet lightweight sheets. These fibrous structures are valued for their remarkable emergent properties, such as a low volume fraction combined with a high mechanical strength~\cite{picu_mechanics_2011}.
However, understanding how the microscopic features of the individual fibers relate to the macroscopic properties of the random fiber network remains an ongoing challenge~\cite{philipse_random_1996,ekman_contact_2014,picu_mechanics_2011}. 
A notable study of fiber packing is the aegagropilae (or sea ball)~\cite{verhille_structure_2017}. These dense balls are formed by entangled fibrous remains of \textit{Posidonia oceanica}, which accumulate and deposit on Mediterranean seashores~\cite{bonnier_revue_1893}.

The main goal of our research is to understand the formation of aggregates in a turbulent flow. 
We expect this process to result from a dynamic equilibrium between the addition of new fibers and the fragmentation/erosion of aggregates. 
To quantify this last process, we need to characterize the specific structural and mechanical properties of aggregates formed in turbulent flows and establish links with existing literature on fiber networks.
In this study, dense spherical aggregates of nylon fibers are formed in a turbulent von Kármán  flow.
Their structural an mechanical properties are characterized and compared with those of natural aegagropilae.
The content of this paper is as follows. 
In section~\ref{SecExpSetup}, we present the experimental setup leading to the formation of fiber aggregates.
In section~\ref{SecTomo}, X-Ray tomographic images are used to characterize some of the structural properties of the aggregate. Then, in section~\ref{SecElasticity}, the effective mechanical properties of the aggregates is probed through mechanical indentation tests.

\section{\label{SecExpSetup}Experimental setup}

\begin{figure}[htbp]
\includegraphics[width=\textwidth]{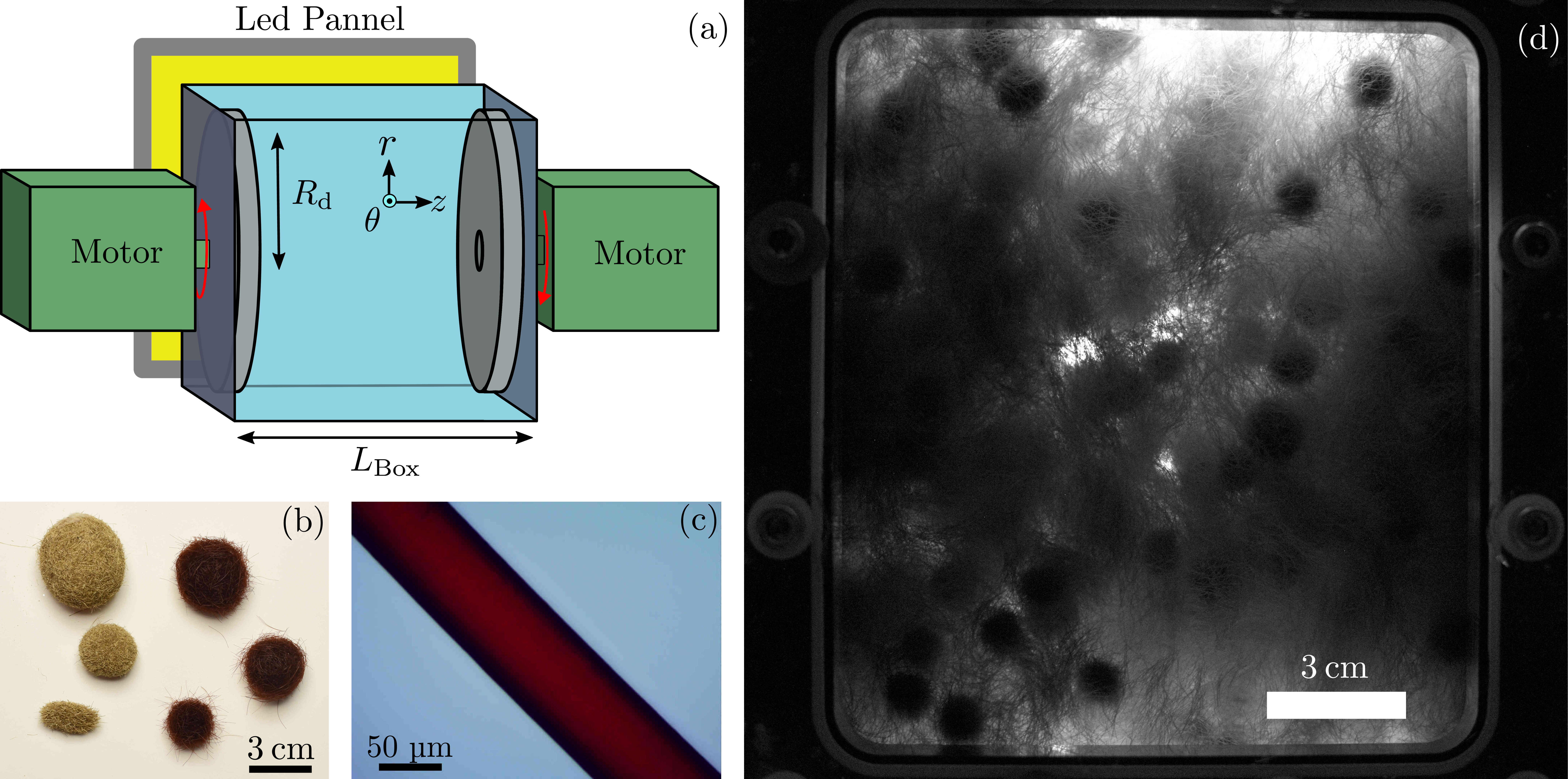}
   \caption{(a) Diagram of the experimental setup. (b) Example of three aggregates obtained in the experiment (three dark red balls on the right side on the image), compared with posidonia balls (three light brown balls of the left side of the image). The black bar measures \SI{3}{cm}. (c) Image of a fiber taken with an optical microscope with a $\times 40$ magnification. (d) Snapshot of a running experiment.}
    \label{Setup}
\end{figure}

Nylon fiber aggregates are formed in the laboratory using a von Kármán turbulent flow, as illustrated in Figure~\ref{Setup}~(a).
The experimental setup consists of a cubic tank filled with water, driven by two counter-rotating smooth disks. 
The side of the tank is $L_\text{Box} = \SI{20}{cm}$ and the radius of the disks are $R_\text{d} = \SI{8}{cm}$.
The flow is mainly driven by two contra-rotating mean flows associated with axial recirculation.
The flow is fully turbulent in the central part of the tank~\cite{machicoane_dynamique_2013}.
The main flow characteristics, measured using PIV without fibers, are summarized in Table~\ref{TableFlow}.

\begin{table}[hbtp]% add [H] placement to break table across pages
    \caption{\label{TableFlow} Flow and fibers main characteristics}
    \begin{ruledtabular}
    \begin{tabular}{cc}
        Motor frequency $f$ & \qtyrange[range-units=single,range-phrase=~--~]{2.5}{21.5}{\Hz} \\
        Root mean squared velocity $U_\text{rms}$ & \qtyrange[range-units=single,range-phrase=~--~]{4.2e-2}{3.8e-1}{m.s^{-1}} \\
        Dissipation rate $\epsilon$ & \qtyrange[range-units=single,range-phrase=~--~]{1e-4}{2e-1}{m^2.s^{-3}} \\
        Kolmogorov length $\eta = \qty(\frac{\nu^3}{\epsilon})^{1/4}$ & \qtyrange[range-units=single,range-phrase=~--~]{4.7e-2}{3.2e-1}{\mm} \\
        Reynolds number at Taylor scale $\Rey = \frac{U_\text{rms} \lambda}{\nu}$ & \qtyrange[range-units=single,range-phrase=~--~]{6.8e2}{1.2e3}{} \\
        Integral scale $L_\text{i}$ & \SI{2}{\cm}\\
        \hline
        fiber length $L$ & \SI{1 \pm 0.02}{\cm}\\
        fiber diameter $d$ & \SI{75 \pm 2}{\micro m}\\
        fiber density $\rho_p$ &\SI{1140 \pm 10}{kg.m^{-3}}\\
        fiber Young Modulus $E_\text{f}$ & \SI{2.9e9}{Pa}
    \end{tabular}
    \end{ruledtabular}
\end{table}

The main fiber properties can be found in Table~\ref{TableFlow} and a microscope view of a fiber is shown in Figure~\ref{Setup}~(c). 
These flexible fibers have a well-defined diameter of \SI{75}{\micro m} and a length \SI{1}{cm} (aspect ratio of \num{133.3}). They possess a smooth surface texture.
A couple of million of these fibers are introduced in the tank prior to the start of the motors, selecting a volume fraction of $\phi_\text{Bulk} \sim \SI{1e-2}{}$.
Then, the motors are operated at a high frequency (\SI{25}{Hz} for \SI{10}{min}) to ensure that no pre-existing aggregates remain in the system. 
After this initial phase, the rotation speed is reduced to the target frequency used in the study. 
Specifically, the results presented in this article correspond to two tested disk rotation frequencies: \qtylist{7;9}{Hz}.

The experiments are conducted over several hours, with durations ranging from \qtyrange{1}{48}{h}, during which aggregates start to form.
This process is illustrated in Figure~\ref{Setup}~(d). 
At the end of the experiment, the aggregates are manually collected and subsequently dried for further analysis.
The number of fibers contained in an aggregate ranges form $10^3$ to $10^4$.
The concentration of the suspension, the fiber dimensions, and the turbulence intensity result in a wide range of aggregate properties. 
Their numbers range from no aggregates to about \num{500} aggregates and their sizes from \qtyrange{0.7}{2}{cm} in radius.
The relationship between the input parameters and the formation of the aggregates as well as their output properties will be addressed in another article.
Here we will only focus on the structural and mechanical properties of the formed aggregates.

An example of the aggregates obtained are presented in Figure~\ref{Setup}~(b) in dark red.
For comparison, Posidonia sea balls are also presented in light brown in Figure~\ref{Setup}~(b). 
Our aggregates are considered spherical, as the mean aspect ratio is of the order of \num{0.93 \pm 0.01}{}.
The aggregates are dense and cohesive. 
They easily maintain their shape under small perturbations such as finger pressure.
While the posidonia sea balls are also dense, cohesive and can be spherical, they often exhibit a preferred axis, leading to an aspect ratio of approximately \SI{0.5}{} \cite{verhille_structure_2017}. 
This difference might result from their formation near the sea floor in nature whereas in the lab experiment, the fibers are homogeneously dispersed in the flow.

To better understand how aggregates maintain their coherence, we characterize the structure of the fiber network using X-Ray tomography.

\section{\label{SecTomo}X-Ray tomography}

To characterize the fiber network of an aggregate, X-ray tomography is used with an RX Solutions EasyTom XL 150 "Mechanic Ultra" microtomograph.
This system is equipped with a sealed micro-focus RX tube operating at \SI{150}{kV} and a \numproduct{1920 x 1536} pixel resolution.
The final resolution achieved ranges from \qtyrange{9.8}{18.5}{\micro\meter}, depending on the size of the aggregate.

\begin{figure}[hbtp]
    \includegraphics[width=\textwidth]{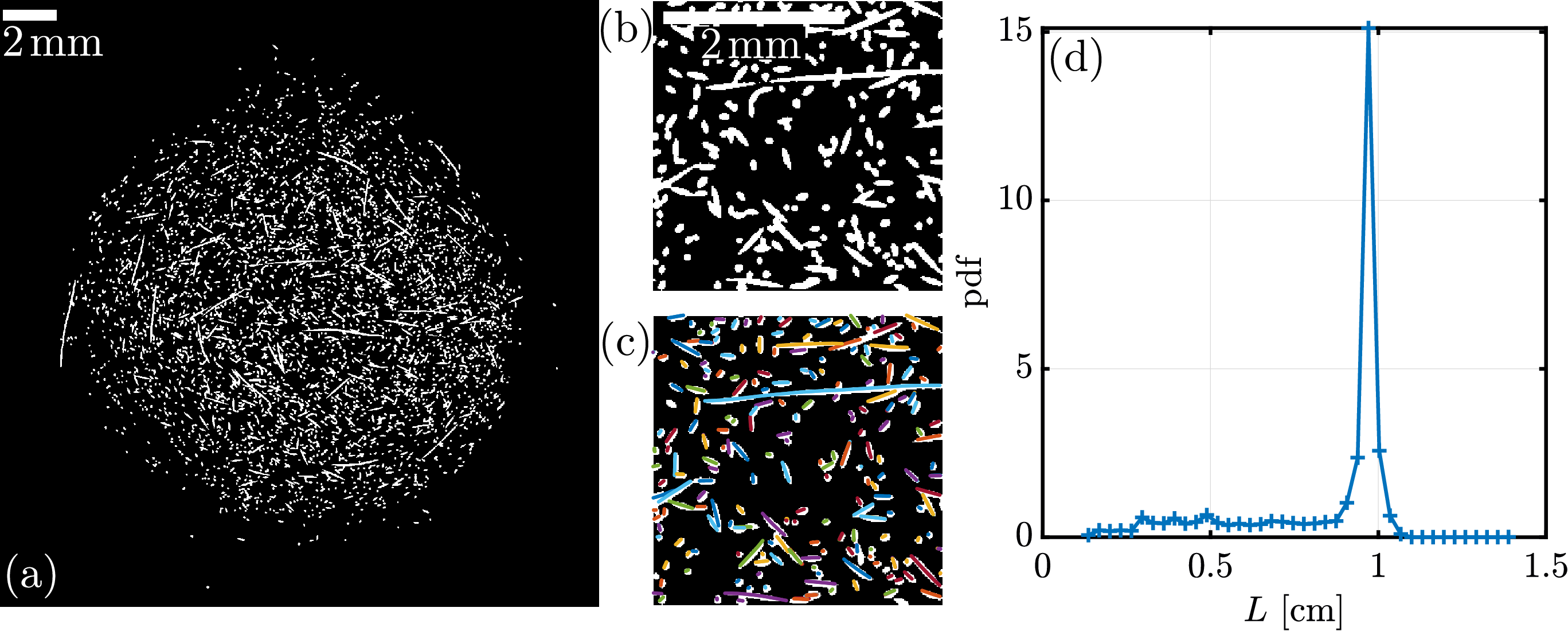}
   \caption{(a) Binarized data of the tomographic reconstruction. (b) corresponds to a zoom of the central region of the image. (c) is the same image on which we added the fit of the fibers separated by colours. (d) Measure of the length of the fibers after reconstruction.}
    \label{FitEtLongueur}
\end{figure}

Following an initial reconstruction using the RX software and binarization, cross-sectional slices of the aggregate can be obtained, as shown in Figures~\ref{FitEtLongueur}~(a) and (b). 
In these images, the fibers are well-resolved, with approximately six voxels across their diameter. 
However, the exact contact points between fibers are ambiguous, complicating their separation.
To address this difficulty, we implemented a custom algorithm to isolate individual fibers. The process is as follows:
i) \textit{Skeletonization:} The 3D volume of the aggregate is skeletonized to identify the fiber extremities.
ii) \textit{Distance Transform:} A distance transform is applied to the inverted binarized volume, enhancing the central regions of the fibers.
iii) \textit{Cylindrical Fitting:} Starting from each extremity, small cylinders are iteratively fitted along the distance-transformed volume. 
The fitting process is constrained to ensure continuity and forward progression, stopping when the fitted cylinder deviates from the fiber. 
The cylinder length is set to $L/30$, which is sufficiently long to follow the fiber direction while separating contact points, yet short enough to maintain local accuracy.
The cylinders have an overlapping of $L/60$. 
Indeed, as discussed in the section~\ref{Coord}, the mean number of contacts per fiber is of approximately $\expval{C} \sim 22$. 
iv) \textit{Merging:} fibers detected from opposite extremities are merged to avoid double-counting. In the end, the fibers are determined by 120 3D coordinates along their length. 

The result of this procedure is shown in Figure~\ref{FitEtLongueur}~(c). The algorithm successfully separates fibers, even in cases of contact. 
However, some fibers are not detected (around \SI{4}{\%}), and occasional duplications occur due to imperfect merging.

Despite these limitations, the histogram of fiber lengths shown in Figure~\ref{FitEtLongueur}~(d) accurately captures the fiber length of around \SI{1}{cm}, even though the algorithm does not impose a fixed length during the fitting process.
The slight difference between the expected fiber length, \SI{1}{cm}, and the measurement, might be explained by the fact that the error in determining the extremities is of the order of $L/16 = \SI{0.65}{cm}$.

Using these tomographic images, we will first characterize the density distribution within the aggregates, followed by an analysis of the coordination number of the fibers and their orientation, and finally, an evaluation of the fiber curvature.

\subsection{\label{Density}Density}

\begin{figure}[hbtp]
\includegraphics[width=0.3\textwidth]{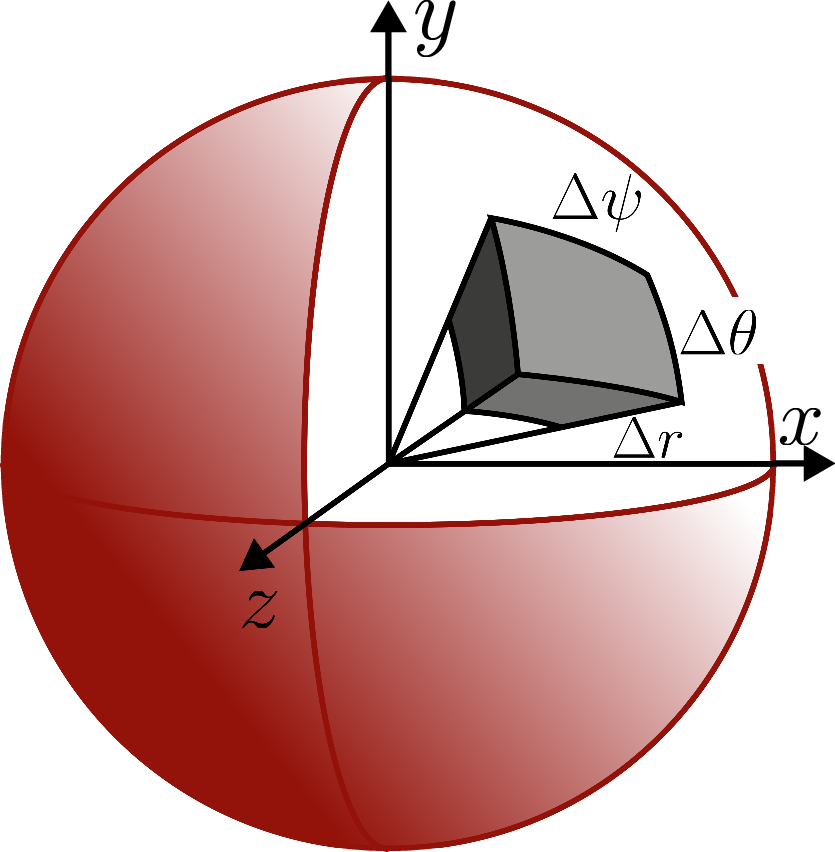}
   \caption{Sketch of the conical sectors taken as volume samples to measure the volume fraction distribution. In our case, $\Delta\theta = \Delta\psi = \pi/4$.
}
    \label{figConical}
\end{figure}

\begin{figure}[hbtp]
\includegraphics[width=\textwidth]{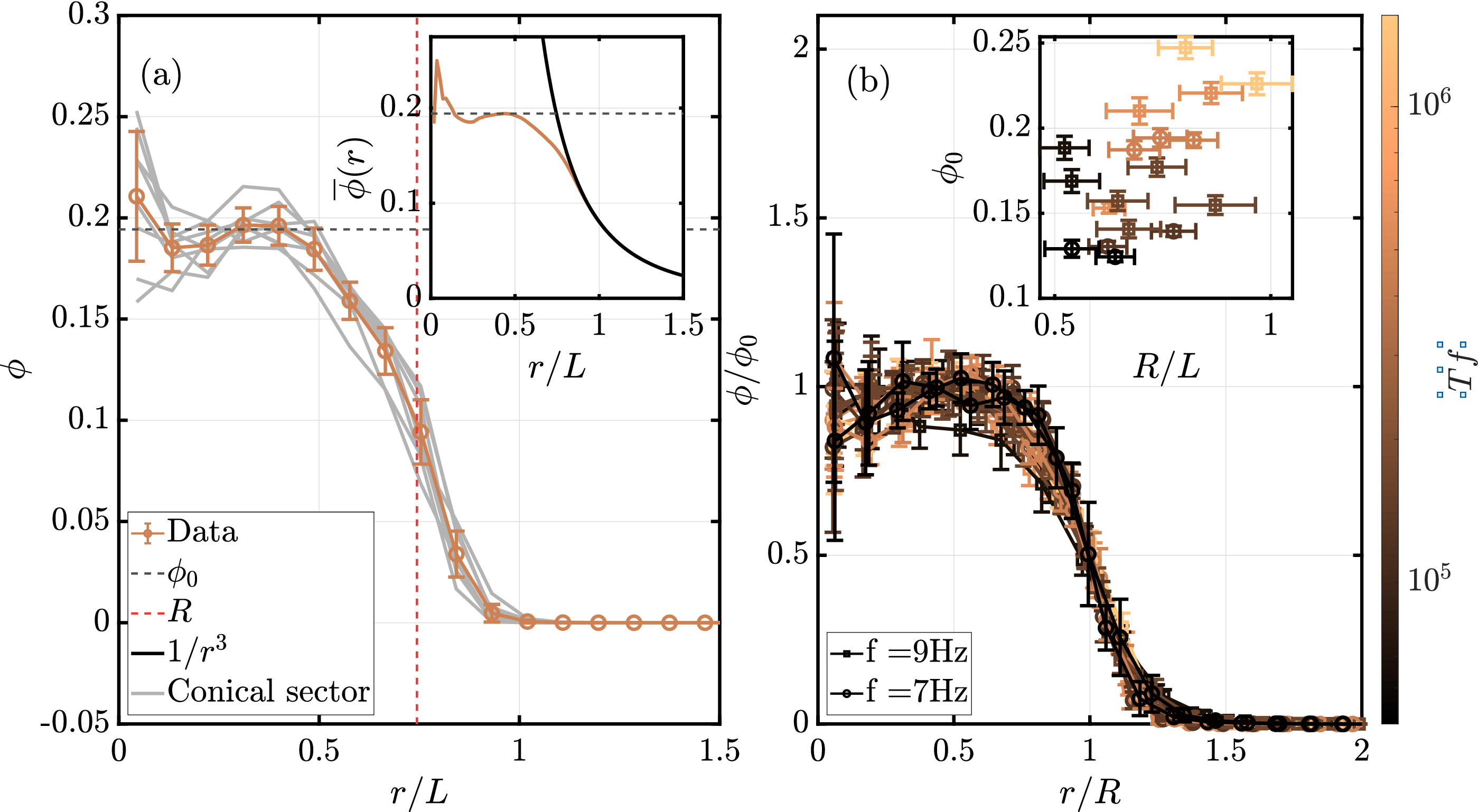}
   \caption{(a) Radial variation of the volume fraction of fibers within an aggregate. The gray curves correspond to the radial variation of the volume fraction restricted to conical sector characterized by a angle $\pi/4$. In the inset, the mean volume fraction $\overline{\phi}(r)$ is plotted as a function of the radius $r$. It corresponds to the volume fraction calculated on a ball of radius $r$. The horizontal dashed line represents $\phi_0$, defined as the maximum of $\overline{\phi}(r)$ for $r/L>0.2$. The black curve is a fit to the latter part of the data in the form $\overline{\phi}(r) = \frac{a}{r^3} +b$. This fit is used to determine the mean radius $R$ of the aggregate, with $a = \phi_0 R^3$. The values of $\phi_0$ and $R$ are indicated on the main graph with black and red dashed lines, respectively.
   (b) Radial evolution of the volume fraction, normalized by the aggregate size and mean volume fraction, for different aggregates. The inset shows the relationship between the mean volume fraction $\phi_0$ and the aggregate radius $R$, as defined in (a). The color bar for both figures corresponds to $Tf$ with $f$ the rotation frequency of the motors and $T$ the duration of the experiment. $Tf$ represents the total number of disk rotations.}
    \label{figDensity}
\end{figure}

The reconstructed data allow us to measure the variation of the volume fraction within the aggregate.  
Indeed, each fiber appears white after reconstruction (see Figure~\ref{FitEtLongueur}), so that the ratio of white volumes to the total volume provides a measure of the volume fraction.  
We first consider subvolumes of the aggregate corresponding to conical sectors, as presented in Figure~\ref{figConical}.  
In our case, $\Delta\theta = \Delta\psi = \pi/4$.  
The result is shown in Figure~\ref{figDensity}~(a) with gray lines.  
We observe that the density is to a good approximation invariant under rotation.

This figure shows that the aggregate maintains an almost constant density up to $r/L \simeq 0.6$, beyond which the density begins to decrease.
The drop is gradual rather than abrupt du to the fuzzy nature of the aggregates, which leads to a reduction in volume fraction at the edges.

Due to these fuzzy edges, the mean volume fraction $\phi_0$ of the aggregate is not easily defined.  
Since the volume fraction is constant in the center of the aggregate, to compute $\phi_0$, we use the average volume fraction up to a given radial distance $r$, represented as $\overline{\phi}(r) = \frac{1}{V(r)}\int_V(r) \phi \, \dd{V}$, with $V(r)$ being the volume of a ball of radius $r$.  
This quantity represents the volume of fibers (white volumes in the tomography) over the volume of a ball of radius $r$.  
This is shown in the inset of Figure~\ref{figDensity}~(a), with $r=0$ corresponding to the center of mass of the aggregate.  
If the volume fraction of the aggregate were a perfect step function (i.e., $\phi(r) = \phi_0$ for all $r \leq R$, the radius of the aggregate, and $0$ for $r \geq R$), $\overline{\phi}(r)$ would be constant and equal to $\phi_0$ for $r \leq R$, and then decreases as $\overline{\phi}(r) = \frac{a}{r^3}$ for $r \geq R$.  
With real data, in the initial region $r/L > 0.2$, the measurement errors are high due to the small control volume.  
We then observe a plateau.  
The value of $\phi_0$ is taken as the maximum of this curve for $r/L > 0.2$.

A fitting function is used on the tail of $\overline{\phi}(r)$ in the form $\overline{\phi}(r) = \frac{a}{r^3} +b$.
The intersection between the fit to the tail of $\overline{\phi}(r)$ and $\phi_0$ is used to determine the mean radius $R$ of the aggregate.

Figure~\ref{figDensity}~(b) shows that the shape of the radial volume fraction profile, $\phi (r)$, is independent of the formation time or motor frequency in the range we tested. 
This behaviour contrasts with observations for Posidonia sea balls, where the volume fraction increases quadratically with the radial position \cite{verhille_structure_2017}.
A possible explanation for this observation could be that Posidonia sea balls are older aggregates, which results in the adhesion of an increasing number of fibers over time. 
Additionally, as the aggregate ages, it undergoes more collisions with the surrounding walls or seabed, resulting in a densification of the structure.

In the inset of Figure~\ref{figDensity}~(b), the analysed X-ray images show that both the mean radius and mean volume fraction increase with $Tf$, the number of disk rotations. However, due to limited data and significant natural variability, it is not possible to infer a robust scaling law.

The typical volume fraction of the considered aggregate is around $\phi_0 = \SI{0.17}{}$. 
For comparison, in randomly packed hard rods, the expected volume fraction is given by Toll's law $\phi_\text{rand}(\frac{L}{d} = 133) \simeq 4d/L=0.03$~\cite{toll_packing_1998}.  %for an aspect ratio of 133 \cite{philipse_random_1996}. 
This indicates that the aggregation process leads to significant compaction compared to random rods packing.
The flexibility of the fibers and their inherent curvature are certainly at the origin of this strong compaction.

\subsection{\label{Coord} Coordination Number}

\begin{figure}[hbtp]
\includegraphics[width=0.5\textwidth]{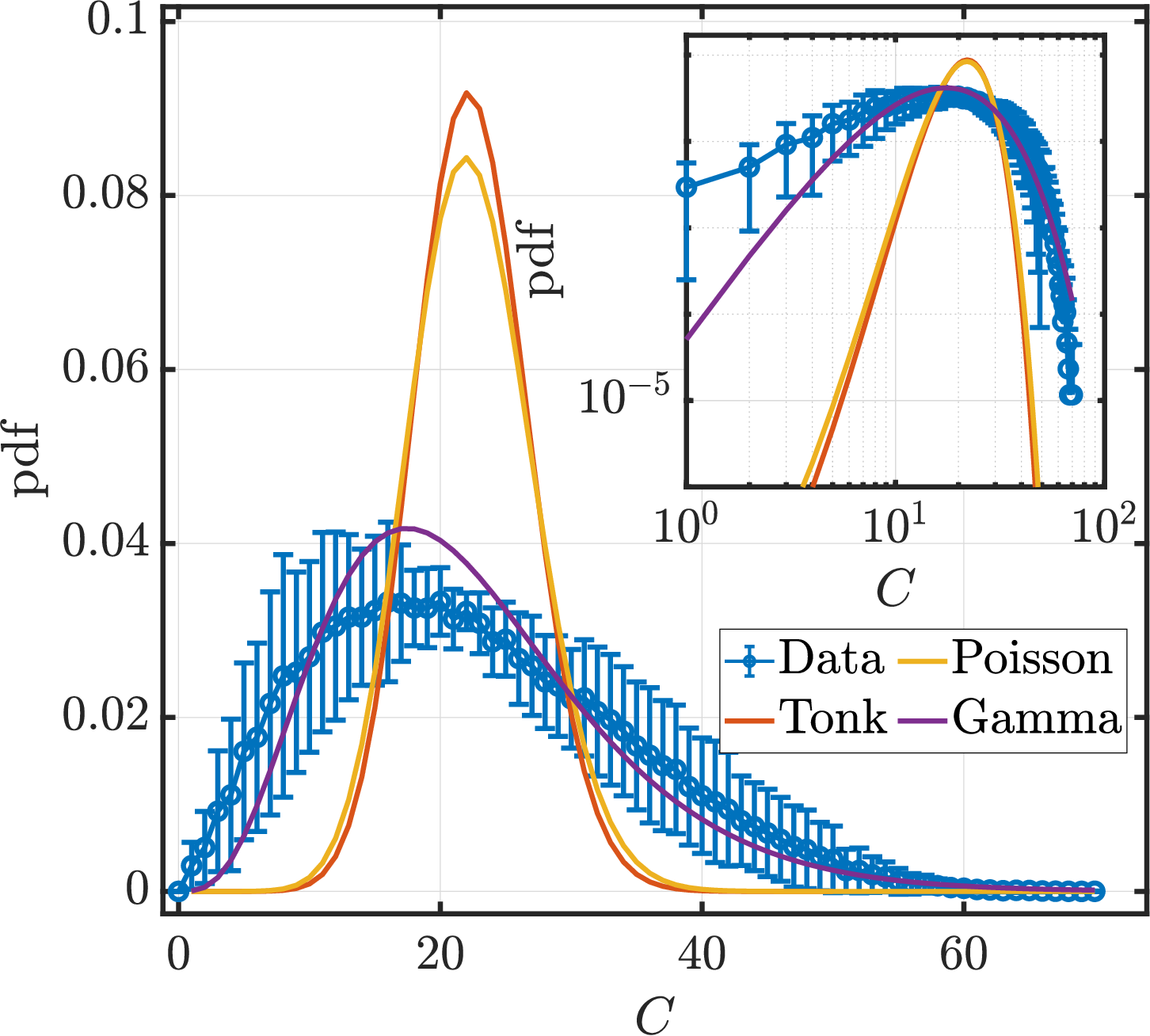}
   \caption{Distribution of the coordination number. The distribution is computed by counting the contacts between the fibers (interpolated with 200 points), considering that there is a contact if the distance between the fibers is less than $s$. We vary $s$ from $d$ to $1.12d$ in 8 steps. The distribution plotted is the mean of the distribution for these different thresholds. The error bars correspond to the standard deviation of each point for the different thresholds. 
   The yellow curve corresponds to the Poisson distribution with parameter $\lambda = \expval{C}$. The red curve corresponds to the two-sided Tonks distribution~\cite{ekman_contact_2014} with parameter $d/L = 133$ and imposing that the mean of the distribution correspond to $\expval{C}$. The purple curve corresponds to a $\Gamma$ distribution with parameters $\alpha = 4.5$ and $\lambda = 0.20$ calculated from the mean and standard deviation of the data.}
    \label{figContact}
\end{figure}

The coordination number corresponds to the number of contacts per fiber.
This quantity is crucial as it influences the stiffness of the network.

For hard rods, the mechanical stability of the random isostatic packing ensures that $4 < \expval{C} < 10$~\cite{blouwolff_coordination_2006}.  
However, in our case, the fibers are curved (Section~\ref{Curv}), and the packing is non-random, as discussed in Section~\ref{Orientation}, therefore, we expect that the coordination number can be greater than 10.

In the case of random hard rod packings, the mean coordination number $\expval{C}$ is given by~\cite{philipse_random_1996}:
\begin{equation} 
\expval{C} \sim 2 \phi \frac{L}{d} \qq{for} \frac{L}{d} \gg 1 
\label{eqc} 
\end{equation}
This estimation is derived from an excluded volume analysis. 
Applying this formula to our system predicts $\expval{C} = 45$.

The reconstruction of fibers allows to measure the coordination number for our aggregate.  
The coordination number is computed by considering the distance between all pairs of fibers.  
Two fibers are considered in contact if this distance is lower than a threshold $s$.  
This threshold is varied from $d$ to $1.12d$.  
Indeed, the fibers are interpolated from the reconstruction using 200 points.  
Therefore, $s = d$ corresponds to the distance expected between the fibers for a contact, while $s \simeq 1.12d$ accounts for cases where fibers are perpendicular and the contact point lies in the middle of the sampled points, corresponding to the maximum expected distance. 
 Figure~\ref{figContact} shows the distribution of the coordination numbers in an aggregate. 
The mean coordination number in our case is $\expval{C} = \num{22.4\pm 4.8}$. 
This discrepancy can be explained by the fact that the measurement is performed on the entire aggregate, where fibers at the edges have fewer contacts than those at the center.
Moreover, we also consider that two fibers can only cross once to avoid infinite connectivity in the case of two parallel fibers, which could also artificially decrease the coordination number.
The preferential orientation of the fibers, as discussed in Section~\ref{Orientation}, can also lead to a decrease in the coordination number.

For hard rods, contacts are typically considered point-like and modeled as successive Bernoulli trials, resulting in a Poisson distribution for the coordination number. This theoretical distribution is shown in Figure~\ref{figContact}, but it does not match our experimental data.

When accounting for the finite size of the contacts due to the relative orientation of the fibers, the Tonks’ law for a one-dimensional gas along a line must be considered~\cite{ekman_contact_2014}:
\begin{equation} 
P_1(C = N) = \frac{\mu L \qty[\mu d_\text{e} \qty(\frac{L}{d_\text{e}} - N + 1)]^N}{\expval{C}_1 N!} \exp\qty(-\mu d\text{e} \qty(\frac{L}{d_\text{e}} - N + 1)) 
\end{equation}
Here, $\mu = \expval{C}_1/\qty(L - \expval{C}_1 d_\text{e})$, $d_\text{e}$ is the apparent size of the contacts, and $\expval{C}_1$ is the mean number of contacts on one side of a fiber.

To account for contacts occurring on all sides of the fibers, Ekman~\cite{ekman_contact_2014} used the convolution of the one-dimensional contact distributions: $P_n(C = N) = P_1 \ast P_n(C = N)$. 
This law is also shown in Figure~\ref{figContact}.
As expected, incorporating the finite size of the contacts narrows the distribution compared to the Poissonian model.
However, it still does not fit our experimental data.

It appears that our data corresponds to a $\Gamma$ distribution:
\begin{equation}
    P_\Gamma(C =N) = \frac{\lambda^\alpha}{\Gamma(\alpha)}N^{\alpha-1}\exp(-\lambda N) 
\end{equation}
The $\Gamma$ distribution, characterized by parameters $\alpha$ and $\lambda$, can be interpreted as the sum of $\alpha$ exponential distributions with parameter $\lambda$. 
In our case, this corresponds approximately to four exponential distributions with $\lambda = 0.17$.

\subsection{\label{Orientation} Orientation}

\begin{figure}[hbtp]
\includegraphics[width=0.8\textwidth]{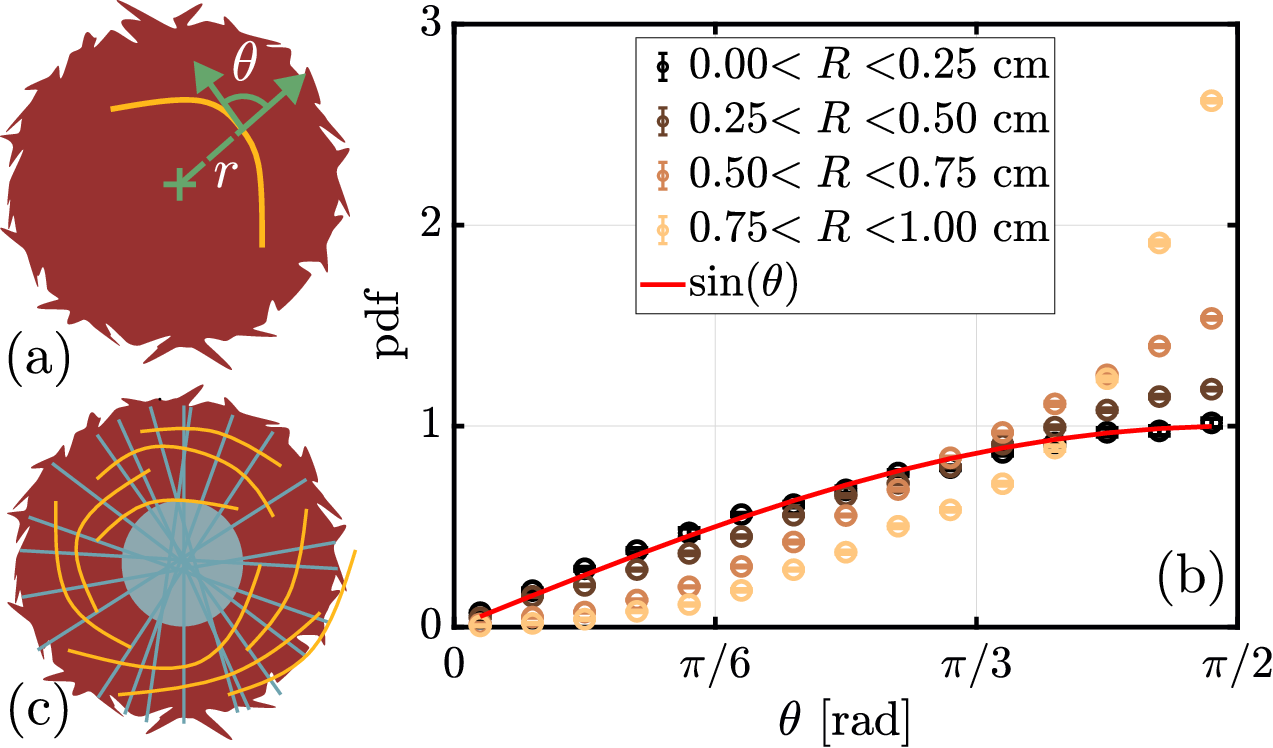}
   \caption{(a) Definition of the angle $\theta$ used to characterize the angular distribution of fibers. (b) Distribution of the angle $\theta$ for different shells of the aggregate. The error bars correspond to the standard deviation obtained by taking 10 times a random sampling of 80\% of the data. In red, the distribution $\sin(\theta)$ corresponding to randomly oriented fibers. (c) Schematic representation of the organisation of the fibers in the aggregate: The blue lines correspond to fibers passing in the core of the aggregate and randomly orientated. The orange lines correspond to fibers wrapped around the aggregate.}
    \label{AllAngle}
\end{figure}

To characterize the angular organization of the aggregate, we define $\theta$ as the angle between the radial direction of a point on a fiber and the tangent to the fiber at that point. 
This is illustrated in Figure~\ref{AllAngle}~(a). 
The angle $\theta(r)$ is calculated from the coordinates of the parametrized fibers, $X(s)$, using the centre of the aggregate as the origin :
\begin{equation}
    \theta = \frac{\va{X}(s+\delta s)-\va{X}(s-\delta s)}{\norm{\va{X}(s+\delta s)-\va{X}(s-\delta s)}}\cdot \frac{\va{X}(s)}{\norm{\va{X}(s)}}
\end{equation}
and $\delta s$ is fixed $\delta s = \SI{1}{mm}$.

Figure~\ref{AllAngle}~(b) shows that fiber orientations depend on their position within the aggregate. 
In the core of the aggregate, the fibers are randomly distributed, following a $\sin(\theta)$ distribution (as expected for random orientations due to the solid angle in spherical coordinates). 
However, in the outer shells, the distribution of $\theta$ deviates from the random case, showing a higher proportion of orthoradial fibers.

This observation suggests that fibers are randomly distributed in the core, but new fibers added during the aggregation process, wrap around the aggregate. 
A similar phenomenon has been observed in Posidonia sea balls, where the outer shell also exhibits an excess of orthoradial fibers \cite{verhille_structure_2017}. 
However, this trend is much more pronounced in our laboratory aggregates. 
One possible explanation for this difference is the angle detection. 
Indeed, for the Posidonia sea balls the angle is found by fitting a cylinder to the 3D tomographic images.
However, there is no history of the fibers for this fit, making the angle determination around the contact points is less precise.
Thanks to our fiber separation, this randomization of the angle $\theta$ at the contact point is less pronounced.
We have tested to use similar technics on the raw tomographic data of our aggregate, leading to comparable results.

\subsection{\label{Curv}Curvature}

\begin{figure}[hbtp]
\includegraphics[width=0.5\textwidth]{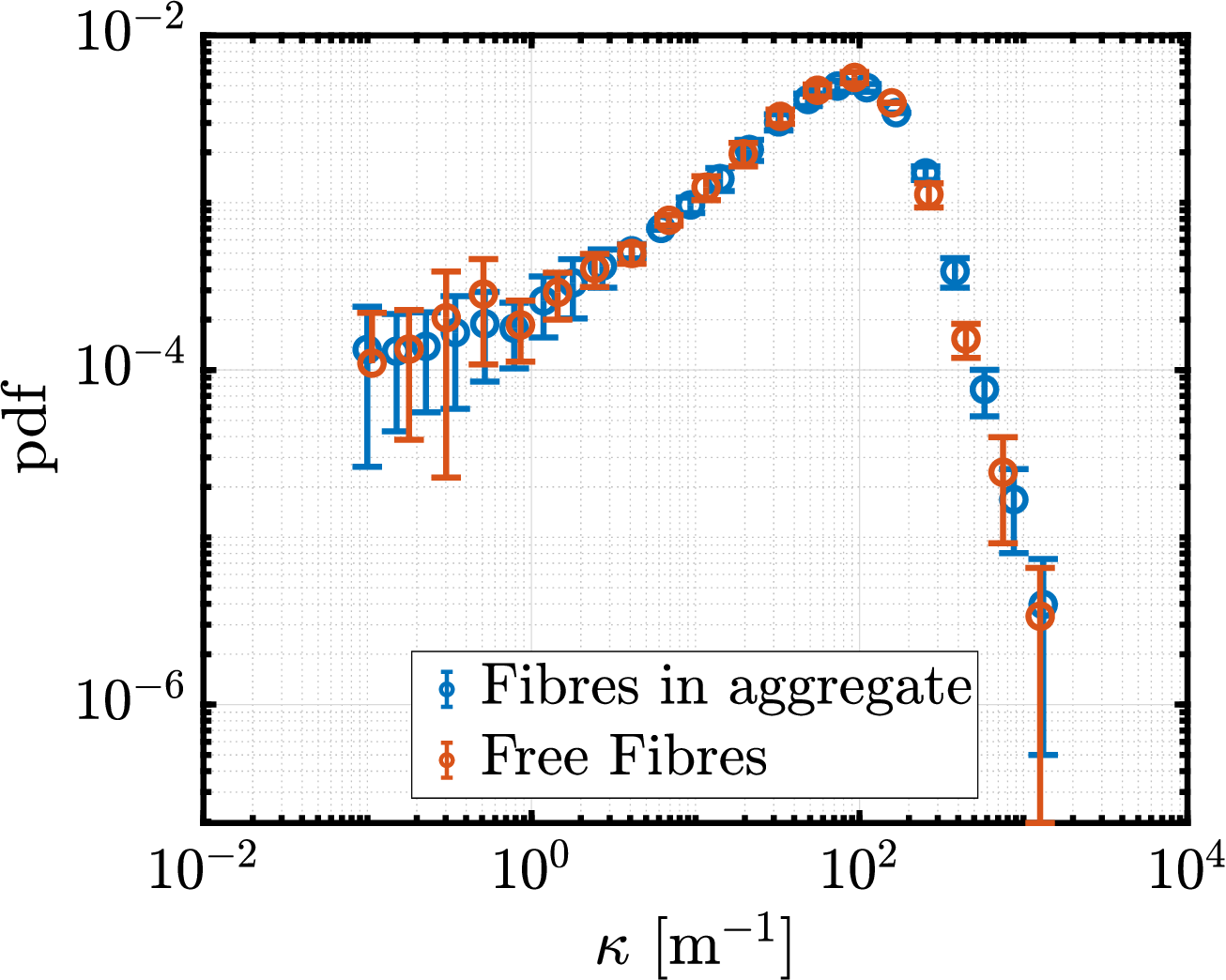}
   \caption{Curvature of the fibers in an aggregate compared with free fibers. The curvature is measured using points separated from a distance $\delta x = L/40$}
    \label{figCourbure}
\end{figure}

The aggregates hold together due to friction between the fibers.
Two main mechanisms can play a role in maintaining normal forces between fibers.
The first one is the hindrance of the fiber rotation and translation in the %entagled
network enhanced by the curvature of the fibers~\cite{gravish_entangled_2016,switzer_flocculation_2004}. 
On Figure~\ref{figCourbure}, we observe that the free fibers are curved at rest with a peak around  $\kappa = \SI{1e2}{m^{-1}}$. 
This curvature, likely due to the production process of the fibers, increases the stability of the aggregates.
The second mechanism is known as elastic interlocking~\cite{soszynski_elastic_1988}.
In that scenario, deformed fibers are pushed against each other at the contact points. 
This elastic force prevents the motion of the fibers due to friction.
In our case, the aggregates are formed by and within turbulence, and we can estimate the curvature induced by the flow on the fibers.
Indeed, according to Brouzet et al.~\cite{brouzet_laboratory_2021}, for fibers with length $L\gg \eta$, the Kolmogorov length (which is the case here), $\kappa l_\text{e} \simeq \beta \qty(L/l_\text{e})^3$, where $\beta$ is a numerical constant, and $l_\text{e} = \qty(EI)^{1/4}/{\qty(\eta\rho\epsilon)^{1/8}}$ is the elastic length of the fiber in turbulence ($\eta$ and $\rho$ are the dynamic viscosity and the density of the fluid, respectively). 
This holds in the limit $L\lesssim l_\text{e}$.
In our case $l_\text{e} = \SI{1.14}{\centi \meter}$.
From the data of Brouzet et al.\cite{brouzet_laboratory_2021} $\beta \simeq \num{5e-2}$.
With this model, the expected curvature of the fibers due to turbulence is of the order of  $\kappa_\text{h} \simeq \SI{1}{m^{-1}}$.
First we note that this value is smaller than the observed natural curvature of the fibers. 
Also, the normalized curvature due to turbulence $\kappa_\text{h} L$ is of the order of $\kappa_\text{h} L \simeq \num{1e-2}\ll 1$ and we might consider neglecting this factor.
However, if we consider a fiber initially straight then bent with a constant curvature $\kappa_ \text{h}$ then the deflection of the fiber is of the order of  $\delta \sim \SI{5e-5}{m}\sim d$ due to the high aspect ratio of the fibers. 
This might help the cohesion process.
However, since the curvature is determined by a second-order derivative, its determination is difficult and our resolution around $\kappa\sim \SI{1}{m^{-1}}$ is not enough to discriminate this effect from natural free fiber curvature. 
Indeed, in Figure~\ref{figCourbure}, the curvatures of both the free and entangled fibers overlap.

In the following section, the structural properties of the aggregates will be linked to their mechanical properties.

\section{\label{SecElasticity} Mechanical properties}
In this section we aim to characterize the effective Young modulus of the aggregates through mechanical indentation tests.

\begin{figure}[hbtp]
\includegraphics[width=0.5\textwidth]{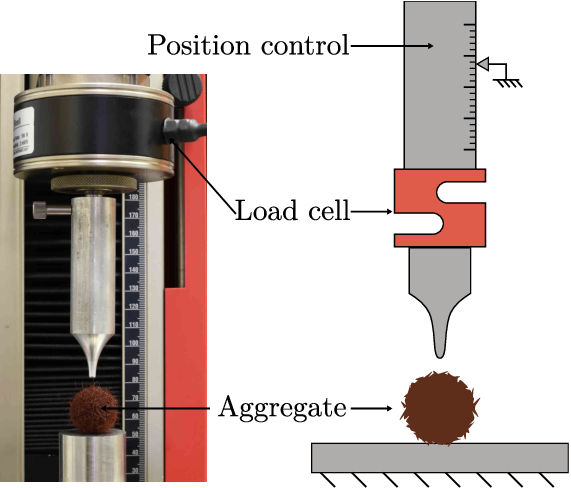}
   \caption{Image and scheme of the experimental setup for the measure of the elasticity of the aggregate. The loading cell is a XForce P with a maximal measuring force of \SI{100}{N}}
    \label{FigElasticite}
\end{figure}

The experimental setup illustrated in Figure~\ref{FigElasticite} allows the characterization of the mechanical response of the aggregate to external forces. 
A ZwickRoell testing machine is used to indent a metal sphere with a radius of $R_\text{ind} = \SI{1.5}{mm}$ into the aggregate at a constant velocity of $v_\text{ind} = \SI{5}{mm/min}$.

\begin{figure}[hbtp]
    \includegraphics[width=\textwidth]{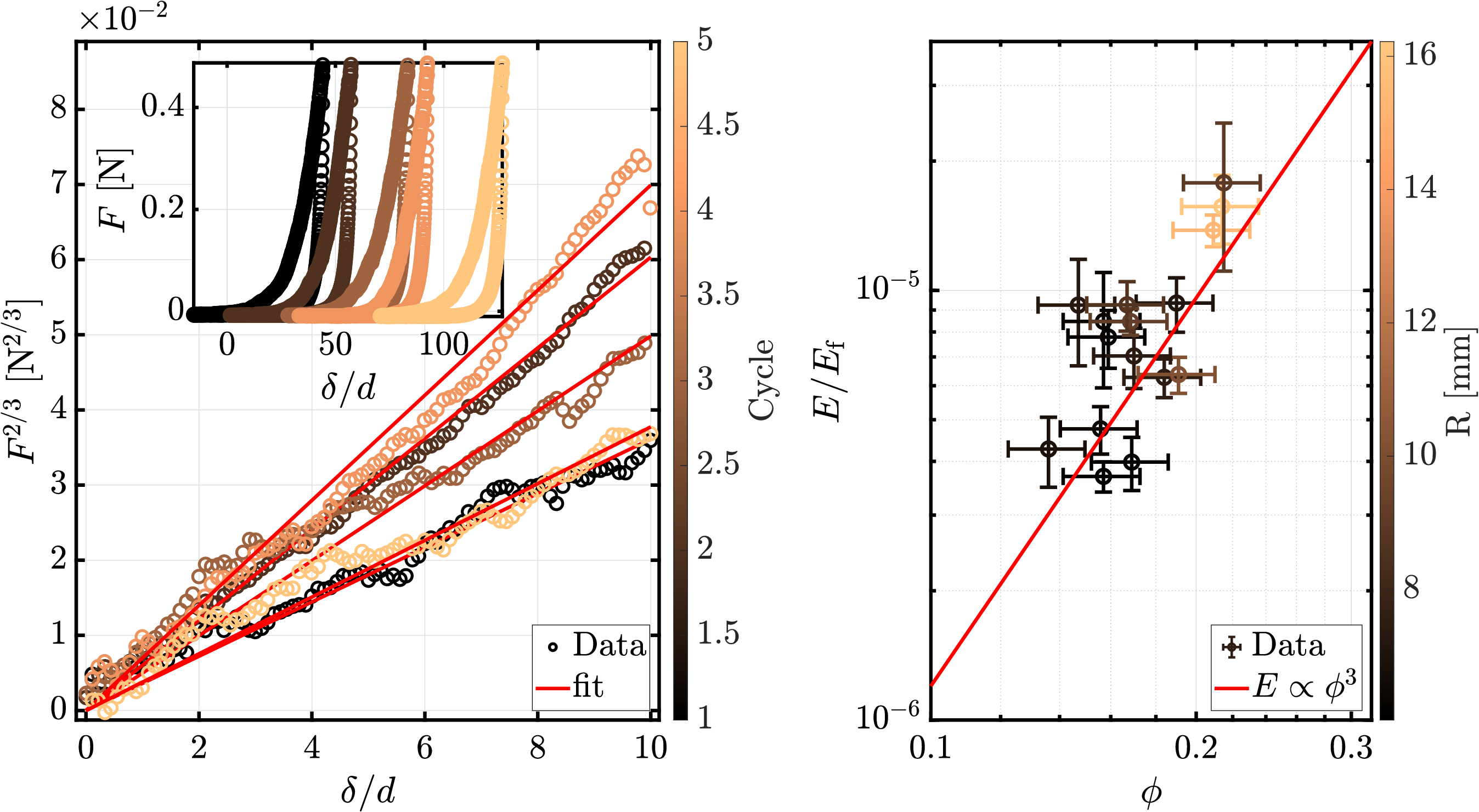}
   \caption{(a) Evolution of $F^{2/3}$ the force to the power $2/3$ as a function of deformation, with a fit based on Hertz's law: $F^{2/3} = b\delta$. In the inset, five typical loading and unloading cycles are presented. Between each cycle, the $\delta/d = 0$ is redefined, and the cycles are shifted manually to increase readability in the inset. The arrows indicate the direction of the hysteresis loop. (b) Evolution of the Young's modulus of the aggregate normalized by the Young's modulus of nylon fibers (tabulated as \SI{2.9}{GPa}) as a function of the aggregate volume fraction. The colorbar corresponds to the aggregate size. Most of the aggregates have a similar size, so that the trend is not expected to be due to size effects. The error bars are estimated to represent 10\% of the values for both $\phi$ and the standard deviation of the Young modulus. The red line corresponds to the fit obtained for Posidonia sea balls~\cite{verhille_structure_2017} : $E\sim \phi^3$, which agrees with our data quite well too.}
    \label{Deformation}
\end{figure}

A load cell measures the force applied to the aggregate as a function of the indentation displacement. 
The indenter size is small relative to the aggregate radius ($R \sim \SI{7}{mm}$) to ensure local measurements that are independent of the aggregate size, yet large compared to the mean spacing between fibers ($l \sim \frac{L}{\expval{C}} \sim \SI{0.5}{mm}$). 
As a first approximation, we consider that the radius of the indenter is large enough to consider the aggregate as a continuous medium where Hertz’s law applies (described later). 
Note that, ideally, this would require $R_\text{ind} > 12l$~\cite{merson_probing_2023}, which is not possible in our case to ensure local measurement. 
An example of five loading cycles is presented in the inset of Figure~\ref{Deformation}~(a).
Between each cycle, the aggregates are rotated randomly. 
The aggregates are not perfect spheres, and therefore, for each indentation, the $\delta/d = 0$ is redefined. 
For the studied aggregates, the maximum indentation force is set to $F_\text{max} = \SI{0.5}{N}$ to avoid drastic reorganization of the fibers between cycles.
Nevertheless, a hysteresis loop is observed, indicating energy dissipation during the process.
Despite this, each curve exhibits a consistent trend, suggesting that the initial indentation phase remains comparable across cycles.

To characterize the equivalent Young’s modulus of the aggregates, we focus on the first part of each curve as an example is shown on Figure~\ref{Deformation}.
For small displacement, the response of the aggregate under loading is expected to follow the laws of linear elasticity, hence, for this planar-spheric configuration to follow the Hertz’s law~\cite{lifshitz_copyright_1986}: $F = \frac{4}{3}ER_\text{ind}^{1/2}\delta^{3/2}$.
Indeed, for a displacement of $\delta\ll R_\text{ind}$ the strain is of the order of $\epsilon \sim \sqrt{\delta/2R_\text{ind}}\simeq 0.4 $ for $\delta= 7 d = \SI{0.52}{mm}$.
One curve gives us a measure of the Young’s modulus of the aggregate, however, this measurement is often noisy due to the intrinsic variability of the aggregates. 
Therefore the effective Young’s modulus is averaged over a repetition of ten measurements while rotating the aggregate between each measurement to probe an undisturbed network at each time.
This allows us to measure the mean effective Young’s modulus of the aggregate.
We note that in nanoindentation, the beginning of the unloading curve is often used to characterize the Young’s modulus of the aggregates~\cite{pharr_measurement_1992}.
However we observed that the value of the Young's modulus estimated with this technique depends on the final depth of the indenter, probably due to a local compaction of the fiber network. 
As for our application we are interested in stiffness of the raw aggregate, we decided to use the beginning of the loading curve, as was done for the Posidonia sea balls~\cite{verhille_structure_2017}. 
This choice is made because the Young’s modulus measured along the unloading branch corresponds to that of a compacted network (with an unknown volume fraction estimated around 0.65) rather than that of the initial uncompacted aggregate (with a known $\phi_0$)

We note that in the literature, several regimes of variation of the Young's modulus have been identified for athermal networks. The more general expression linking the stress $\sigma$ to the volume fraction $\phi$ is given by:  
\begin{equation}
\sigma \propto \qty(\phi^n - \phi_0^n)
\end{equation}
where $\phi_0$ is the volume fraction at the percolation limit, and $n$ is an exponent depending on the compaction.  
At low compaction, van Wyk's derivation~\cite{van_wyk_20note_1946} leads to $n = 3$.  
At high compaction, the fibers are preferentially oriented perpendicular to the direction of indentation, and $n$ is found to be around $n = 5$ according to Toll’s law~\cite{toll_packing_1998}.  
Between these two regimes, Picu et al.~\cite{picu_compression_2022} find a value of $n = 2$.  
For more informations about these different scalings, the reader should refer to the review~\cite{picu_mechanics_2011}  or the book~\cite{picu_network_2022} by R.C. Picu.

For Posidonia sea balls, the effective Young’s modulus has been modeled~\cite{verhille_structure_2017} to account for the fibrous properties of the aggregate, following the van Wyk model~\cite{van_wyk_20note_1946}, therefore with $n=3$. 
In our case, only the scaling in $\phi$ was taken into account.
Indeed, we expect $\phi\gg\phi_0$ as $\phi\simeq 0.17$ is greater than the random packing case ($\phi_\text{rand} \simeq =0.03$~\cite{toll_packing_1998}) and therfore we expect $\sigma \propto \phi^n$ to hold with a good approximation.

This model is derived by equating the bending energy of the fibers to the bending energy of a deformed sphere.
Similarly, the derivation of this model can be expressed in terms of forces:
For a homogeneous body characterized by a Young’s modulus $E$, the force associated with a displacement $\delta$ is given by $F_\text{h} = \frac{4}{3}E R_\text{ind}^{1/2}\delta^{3/2}$.
This force must be identified to the force acting on the fibrous medium.
The force on a single fiber can be expressed as $   F_\text{f0} \sim E_\text{f} I L{\kappa}/{w^2}$,
where $E_\text{f}$ is the Young’s modulus of a fiber, $I = \pi d^4/64$ is the moment of inertia of the fiber, and $w \sim \qty(R_\text{ind} \delta)^{1/2}$~\cite{lifshitz_copyright_1986} is the typical transverse distance over which bending occurs.
Here, $\kappa \sim \delta / l^2$ represents the curvature of the fiber, with $l = L / \expval{C} \sim d / (2\phi)$~\cite{philipse_random_1996} as the typical distance between contacts.
The total force is then $F_\text{f} \sim F_\text{f0} N_\text{f}$, where $N_\text{f} \sim {4 w^3 \phi}/\qty(\pi d^2 L)$ represents the number of deflected fibers. Identifying these two forces yields
\begin{equation}
    E \sim E_\text{f} \phi^3
    \label{eq:final}
\end{equation}

This result is consistent with our experimental data, as shown in Figure~\ref{Deformation}~(b).
Our derivation can only determine the scaling law, as the prefactors are not considered.

\section{\label{Disc}Conclusion}

In this study, we have demonstrated the ability to create synthetic nylon fiber aggregates within a von Kármán turbulent flow.
Using X-ray tomography, we characterized the density distribution of these aggregates. 
Our findings indicate that the density profile remains unchanged regardless of the formation parameters tested.

The mean volume fraction of the aggregates increases with their sizes as well as with the duration of the experiments.
The observed volume fractions exceed those typically found in the random packing of rods~\cite{philipse_random_1996}.

The coordination number of the fibers follows a $\Gamma$ distribution, with a mean contact number consistent with the scaling laws based on excluded volume argument~\cite{philipse_random_1996}.
Furthermore, the fibers within the aggregates are not randomly oriented; rather, the structure comprises two distinct regions: an inner core of randomly oriented fibers surrounded by an external shell of orthoradial fibers.
With the resolution of our measurements, no significant changes have been found in the distribution of fiber curvatures within the network compared to free fibers. 
Finally, mechanical indentation experiments, interpreted using an elastic sphere model, show that the Young’s modulus of the fiber aggregates depend on the volume fraction as $E \sim E_\text{f} \phi^3$. This experimental result is consistent with previous findings for similar fiber aggregates~\cite{verhille_structure_2017}.

\bibliography{bib.bib}

\end{document}